\input{aipcheck}

\documentclass[
    ,final            
  ]
  {aipproc}

\layoutstyle{6x9}

\begin{document}

\title{The heat flux from a relativistic kinetic equation with a simplified
collision kernel}


\author{A. Sandoval-Villalbazo}{
address={Depto. de Fisica y Matematicas, Universidad Iberoamericana,
Prolongacion Paseo de la Reforma 880, Mexico D. F. 01219, Mexico.}
}

\author{A. L. Garcia-Perciante}{
address={Depto. de Matematicas Aplicadas y Sistemas, Universidad Autonoma
Metropolitana-Cuajimalpa, Artificios 40 Mexico D.F 01120, Mexico.}
}

\author{L.S. Garcia-Colin}{
address={Depto. de Fisica, Universidad Autonoma Metropolitana-Iztapalapa,
Av. Purisima y Michoacan S/N, Mexico D. F. 09340, Mexico. Also at
El Colegio Nacional, Luis Gonzalez Obregon 23, Centro Historico, Mexico
D. F. 06020, Mexico}
}

\begin{abstract}
 We show how using a special relativistic kinetic equation with a BGK-
like collision operator the ensuing expression for the heat flux can
be casted in the form required by Classical Irreversible Thermodynamics.
Indeed, it is linearly related to the temperature and number density
gradients and not to the acceleration as the so-called "first order
in the gradients theories" contend. Here we calculate explicitly the
ensuing transport coefficients and compare them with the results obtained
by other authors.
\end{abstract}

\maketitle


The purpose of this work is to calculate the explicit form for the
heat flux in an ideal relativistic gas to find out what is the form
of the constitutive equation as required by Classical Irreversible
Thermodynamics (CIT), in the temperature and particle number density
representation. This result is important by itself because it will
help in clarifying whether or not such constitutive equation should
contain a term proportional to the acceleration as it has been claimed
by many authors. To accomplish this task, we start by considering
the simplified relativistic Boltzmann equation in the absence of external
forces:

\begin{equation}
v^{\alpha}f_{,\alpha}=J\left(f,\, f^{\prime}\right)\label{eq:1}
\end{equation}

The term on the right hand side of Eq. (\ref{eq:1}) is the collision
operator which is modeled by means of the BGK approximation \cite{CK}.
In Eq. (\ref{eq:1}), the molecular four velocity, $v^{\alpha}$ is

\noindent \begin{equation}
v^{\alpha}=\left[\begin{array}{c}
\gamma w^{\ell}\\
\gamma c\end{array}\right]\label{eq:1.1}\end{equation}
where $w^{\ell}$ is the molecular three-velocity and
$\gamma=\left(1-w^{\ell}w_{\ell}/c^{2}\right)^{-1/2}$
is the usual relativistic factor. All greek indices run from 1 to
4 and the latin ones run up to 3. The number density reads

\noindent \begin{equation}
n=\int f\gamma dv^{*}\label{eq:01}\end{equation}
with $dv^{*}=\gamma^{5}\frac{cd^{3}w}{v^{4}}$ \cite{Liboff}. Now,
standard kinetic theory leads to the energy balance equation \cite{CK,degrootrel}:
\begin{equation}
\frac{\partial}{\partial t}\left(n\left\langle mc^{2}\gamma\right\rangle \right)+\left(n\left\langle mc^{2}\gamma w^{\ell}\right\rangle \right)_{;\ell}=0\label{Baley}\end{equation}
where the thermodynamic average is defined through: \begin{equation}
\left\langle \psi\right\rangle =\frac{1}{n}\int\gamma \psi fdv^{*}\label{prom}\end{equation}
 In Eq. (\ref{Baley}) one can easily identify the heat flux as: \begin{equation}
J_{[Q]}^{\ell}=n\left\langle mc^{2}v^{\ell}\right\rangle =mc^{2}\int v^{\ell}f\gamma dv^{*}\label{Baley2}\end{equation}
In a Chapman-Enskog expansion, the distribution function is written
as \begin{equation}
f=f^{\left(0\right)}\left(1+\phi\right)\label{eq:3}\end{equation}
where $\phi$ is the first order correction in the Knudsen parameter,
a weighted measure of the gradients in the system. For particles of
rest mass $m$, relativistic parameter $z=\frac{kT}{mc^{2}}$ and
in the non-degenerate case, the equililibrium function reads \cite{CK,degrootrel}:

\begin{equation}
f^{\left(0\right)}=\frac{n}{4\pi c^{3}z\mathcal{K}_{2}\left(\frac{1}{z}\right)}e^{\frac{u^{\beta}v_{\beta}}{zc^{2}}}\label{eq:4}\end{equation}
in which $k$ is Boltzmann's constant, $u^\beta$ the hydrodynamic four-velocity and $\mathcal{K}_{2}$ is the
modified Bessel function of the second kind. We remind the reader
that the first order correction $\phi$ contains already the dissipative
effects in the system.

If $\tau$ denotes the collisional time appearing in the BGK model,
using the functional hypothesis \cite{uf}, the function $\phi$ in
Eq. (\ref{eq:3}) is given in terms of the thermodynamic forces as

\begin{equation}
\phi=-\tau v^{\alpha}\left(\frac{\partial f^{\left(0\right)}}{\partial n}n_{,\alpha}+\frac{\partial f^{\left(0\right)}}{\partial T}T_{,\alpha}+\frac{\partial f^{\left(0\right)}}{\partial u^{\beta}}u_{;\alpha}^{\beta}\right)\label{eq5}\end{equation}
where the derivatives of the Jüttner distribution function are \begin{equation}
\frac{\partial f^{\left(0\right)}}{\partial n}=\frac{f^{\left(0\right)}}{n}\label{eqq}\end{equation}
 \begin{equation}
\frac{\partial f^{\left(0\right)}}{\partial T}=\left(-1-\frac{\mathcal{K}_{1}\left(\frac{1}{z}\right)}{2z\mathcal{K}_{2}\left(\frac{1}{z}\right)}-\frac{\mathcal{K}_{3}\left(\frac{1}{z}\right)}{2z\mathcal{K}_{2}\left(\frac{1}{z}\right)}\right)\frac{f^{\left(0\right)}}{T},\label{eq8}\end{equation}
and \begin{equation}
\frac{\partial f^{\left(0\right)}}{\partial u^{\beta}}=\frac{v_{\beta}}{zc^{2}}\ f^{\left(0\right)}\label{eq9}\end{equation}
We now introduce Euler's equations for the time derivatives in (\ref{eq5}). A hydrodynamic
acceleration term appears in the last term of Eq. (\ref{eq5}) for
$\alpha=4$. In the standard approach this term is expressed by the
pressure gradient using Euler's equation. This would lead to an awkward
representation involving $n$, $T$, $p$ and $u^{\alpha}$. However,
in order to comply with the requirements of CIT, using the fact that
the ordinary non-equilibrium state variables of the fluid are $n$,
$u^{\alpha}$ and $T$, we substitute the pressure gradient through
the equation of state $p=nkT$, which readily leads to the following
form for the heat flux
\begin{equation}
J_{[Q]}^{\ell}=-L_{TT}\frac{T^{,\ell}}{T}+L_{nT}\frac{n^{,\ell}}{n}\label{eq:10}
\end{equation}
Equation (\ref{eq:10}) is of the canonical form required by CIT,
the thermodynamic forces are the temperature and the density gradients,
respectively. The transport coefficients are given by
\begin{equation}
L_{T}=\frac{nk^{2}T^{2}}{m}\tau f_{T}\left(z\right)\thinspace \thinspace,\thinspace \thinspace L_{n}=\frac{nk^{2}T^{2}}{m}\tau f_{n}\left(z\right)\label{eq:11}
\end{equation}
where the functions $f_{T}$ and $f_{n}$ are given by\begin{eqnarray}
f_{T}\left(z\right) & = & \left(\frac{1}{z}-\left(4z+\frac{\mathcal{K}_{1}
\left(\frac{1}{z}\right)}{\mathcal{K}_{2}\left(\frac{1}{z}\right)}\right)^{-1}
\right)\left(\frac{1}{z}+5\frac{\mathcal{K}_{3}\left(\frac{1}{z}\right)}{\mathcal{K}_{2}
\left(\frac{1}{z}\right)}\right)\nonumber \\
& - & \left(1+\frac{\mathcal{K}_{1}\left(\frac{1}{z}\right)}{2z\mathcal{K}_{2}
\left(\frac{1}{z}\right)}+\frac{\mathcal{K}_{3}\left(\frac{1}{z}\right)}{2z\mathcal{K}_{2}\left(\frac{1}{z}
\right)}\right)\frac{\mathcal{K}_{3}\left(\frac{1}{z}\right)}{z\mathcal{K}_{2}\left(\frac{1}{z}\right)}\label{eq:12}\end{eqnarray}
\begin{equation}
f_{n}\left(z\right)=\left(4z+\frac{\mathcal{K}_{1}\left(\frac{1}{z}\right)}{\mathcal{K}_{2}\left(\frac{1}{z}
\right)}\right)^{-1}\left(\frac{1}{z}+5\frac{\mathcal{K}_{3}\left(\frac{1}{z}\right)}{\mathcal{K}_{2}
\left(\frac{1}{z}\right)}\right)-\frac{\mathcal{K}_{3}\left(\frac{1}{z}\right)}{z\mathcal{K}_{2}
\left(\frac{1}{z}\right)}\label{eq:13}\end{equation}
and are shown in Figure 1.

\begin{figure}
\begin{centering}
\includegraphics[width=4in,height=2in]{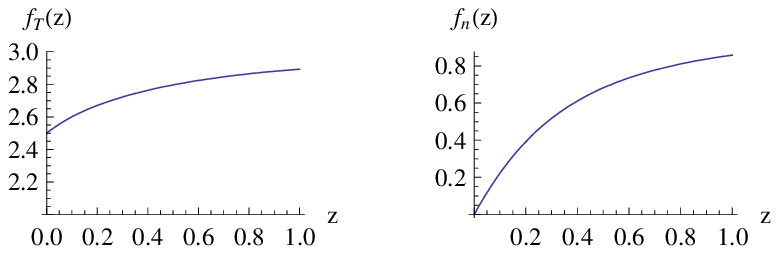}\par
\end{centering}

\caption{The functions $f_{T}\left(z\right)$ (left) and $f_{n}\left(z\right)$
(right).}
\end{figure}

The coefficient given in Eq. (\ref{eq:13}) has not been analyzed
before, although it has been proposed in Ref. \cite{prd}. Its presence
in the constitutive equation for the heat flux given in Eq. (\ref{eq:10}),
which has the structure required by CIT, guarantees the validity of
Onsager's regression of fluctuations hypothesis. Other authors obtain
an expression for the heat flux in terms of the acceleration or pressure
gradient which is, not only against the tenets of CIT, but also gives
rise to the so-called generic instabilities for the relativistic fluid
\cite{GRG,HL}. In contrast, our approach provides a picture in which
fluctuations do not grow exponentially within the first order in the
gradients description of the relativistic simple fluid.

Two important facts shown in Figure 1 are worth pointing out as a
final remark. Firstly, we have obtained an analytical expression for
the thermal conductivity for the relativistic gas, using in the BGK
approximation. This result is new and differs from previous calculations
since this is the first time the representation given in Eq. (\ref{eq:10})
is explicitly used. Notice also how this coefficient yields the correct
non-relativistic limit $L_{T}\sim\frac{5}{2}\frac{nk^{2}T^{2}}{m}\tau$.
Secondly, the figure on the right shows a novel result, namely a coefficient
relating a heat flux to a number density gradient present only in
the relativistic case. This coefficient vanishes for $z\rightarrow0$.



\bibliographystyle{aipproc}   


\end{document}